\documentclass[10pt]{article}
\usepackage[latin1]{inputenc}
\usepackage{amsmath,amssymb,amsthm}
\usepackage{graphicx}
\usepackage{epsfig}

\newtheorem{theo}{Theorem}[section]

\newtheorem{lem}[theo]{Lemma}

\newcounter{listagem}
\newcommand{\blista}{\begin{list}{\roman{listagem})}{\usecounter{listagem}}}
\newcommand{\elista}{\end{list}}

\newcommand{\beq}{\begin{equation}}
\newcommand{\eeq}{\end{equation}}
\newcommand{\beqn}{\begin{eqnarray}}
\newcommand{\eeqn}{\end{eqnarray}}
\newcommand{\ov}{\overline}
\newcommand{\ul}{\underline}

\newcommand{\Cl}{{C \kern -0.1em \ell}}
\newcommand{\BZ}{\mathbb{Z}}
\newcommand{\BR}{\mathbb{R}}
\newcommand{\BC}{\mathbb{C}}
\newcommand{\BN}{\mathbb{N}}
\newcommand{\BH}{\mathbb{H}}
\newcommand{\f}{\mathfrak{f}}
\newcommand{\fm}{\mathfrak{f}^\dagger}

\newcommand{\ds}{\displaystyle}

\begin{document}

\title{Fundamental Solutions of
the Instationary Schr\"odinger Difference Operator}

\author{Paula Cerejeiras ~~~~ Nelson Vieira\\
{\small Department of Mathematics,}\\{\small University of Aveiro,}\\{\small 3810-193 Aveiro, Portugal.} \\
{\small E-mails: pceres@ua.pt, ~nvieira@ua.pt}} \maketitle

\begin{abstract}
In this paper we will study the existence of fundamental solutions
for the explicit and implicit backward time dependent Sch\"odinger
equation, via discrete Fourier transform and its symbol for the
Laplace operator. In both cases we will prove that the discrete
fundamental solutions obtained converges to the continuous
fundamental solution
in the $l_1-$norm sense.\\

\\

\end{abstract}

\section{Introduction}

The potential theory is a useful tool to solve boundary value
problems by the help of integral equations on the boundary. For
constructive analytical considerations and also in the case of
numerical applications it is necessary to have an explicit
expression of the fundamental solution. However, the integral
representations obtained using potential theory are not suitable for
an explicit computation of the solutions, due to unacceptable
convergence rates of the integral's numerical approximations (for
more details see \cite{GS}).

A more acceptable alternative in computational terms is given by the
use of finite difference approximations. In fact, the connection
between the potential theory and finite difference theory is based
in the possibility of obtain explicit expressions for the
fundamental solutions. This second approach has been applied to
elliptic operators, for example, in \cite{DS} and \cite{S}, where
elliptic difference operators were studied and representations for
their fundamental solutions were given. However, this approach is
restricted to the stationary case. In fact, when one takes the time
evolution into account, for example, when the operators are
parabolic, additional difficulties appear. One of the main
difficulties lyes in its fundamental solutions, which must be study
in distributional sense.

This paper is based on the work developed in \cite{GS}, where an
orthogonal decomposition of the function spaces in terms of
subspaces of null-solutions of the correspondent Dirac operator were
given. In order to extend this approach to the nonstationary case we
introduce Witt basis, as done in \cite{CKS}. This will gives us the
possibility of apply techniques of the elliptic function theory to
parabolic domains.

For the study of the continuous Schr\"odinger equation
    \begin{eqnarray}
    (-\Delta_x - i\partial_t)u(x,t) & = & \delta(x,t),
    \label{26*}
    \end{eqnarray}
we need to study the fundamental solution of the difference equation
    \begin{eqnarray*}
    \left( \left( -\Delta_h - i \partial_\tau \right) E_{h,\tau} \right)
    (h\ul{m},\tau k) & = &\delta_{h,\tau}(h\ul{m},\tau k).
    \end{eqnarray*}

For that purpose, two approaches are available to us: either one
considers the explicit equation
    \begin{eqnarray}
    \left( \left( -\Delta_h - i \partial_\tau \right) E_{h,\tau} \right)
    (h\ul{m},\tau k) & = & \left\{ \begin{array}{ccc}
                        \frac{1}{h^3\tau} & \mbox{if} & (h\ul{m},\tau k)=(\ul{0},0) \\
                        & & \\
                        0 & \mbox{if} & (h\ul{m},\tau k) \neq (\ul{0},0)
                        \end{array}\right., \label{*}
    \end{eqnarray}
where $\Delta_h$, $\partial_\tau$ and $\delta_{h,\tau}$ are the
discrete operators of the Laplacian, the discrete time derivation
and the discrete delta function; or we consider the implicit version
of the equation (\ref{*}), with an additional time-step.

The paper will be divided as follows: on Section 2, we present some
basic notions about Clifford analysis and finite diferences
approximations.

On Section 3 we shall study the existence and behavior of
fundamental solutions for the explicit non-stationary Schr\"odinger
equation (\ref{*}); this study will be divided in three steps:
first, we shall prove convergence of the discrete fundamental
solution to the continuous one on a bounded domain. On the second
step, we prove convergence on a domain of type $G_h \times
[T_0,+\infty[,$ $T_0>0$. We will finalize this section with the
proof of convergence for the domain $\BR_h^3 \times
]0,+\infty[_\tau$.

On Section 4 we prove the existence of a discrete fundamental
solution for the implicit Schr\"odinger equation and that this
fundamental solution converges to the continuous one.

In Sections 3 and 4 we will omit a general discussion in discrete
spaces of distributions because, on the one hand we want to
underline the analogy with the continuous case. On the other hand we
are interested in convergence results in norms as strong as
possible. Therefore, we investigate if $E_{h,\tau}$ belongs to the
space $l_1(\BR_h^3 \times \BR_\tau^+)$.

We will present an explicit expression for discrete fundamental
solution of both the explicit and implicit backward Schr\"odinger
equation and we will prove that these solutions belongs to the space
$l_1^{loc}(\BR^3_h \times \BR_\tau^+)$. Also, we will prove that
these fundamental solutions converges to the continuous one in
$l_1-$norm, when the size of the mesh tends to zero.


\section{Preliminaries}

\subsection{Clifford Analysis}

Consider the $n$-dimensional vector space $\BR^n$ endowed with a
standard orthonormal basis $\{ e_1, \cdots, e_n \}$  and satisfying
the multiplication rules $e_i e_j + e_j e_i = -2\delta_{i,j}.$

We define the universal Clifford algebra $\Cl_{0,n}$ as the
$2^{n}$-dimensional associative algebra with basis given by $
e_\emptyset = 1,$ as the scalar unit, and $ e_A = e_{h_1} \cdots
e_{h_k},$ where $A = \{ h_1, \ldots, h_k \} \subset N = \{ 1,
\ldots, n \}$, for $1 \leq h_1 < \cdots < h_k \leq n$. Each element
$x \in \Cl_{0,n}$ will be represented by $x=\sum_{A} x_A e_A$ and
each non-zero vector has their multiplicative inverse given by
$\frac{-x}{|x|^2}.$ We denote the (Clifford) conjugation $x
\rightarrow \ov{x}^{\Cl_{0,n}}$ by means of its action on the basis
elements
    \begin{eqnarray*}
    \ov{1}^{\Cl_{0,n}}=1 \qquad \ov{e_j}^{\Cl_{0,n}}
    = -e_j \qquad \ov{ab}^{\Cl_{0,n}}=\ov{b}^{\Cl_{0,n}}\ov{a}^{\Cl_{0,n}}.
    \end{eqnarray*}

We introduce the complexified Clifford algebra $\Cl_n$ as the
tensorial pro\-duct
    \begin{eqnarray*}
    \mathbb{C} \otimes \Cl_{0,n} = \left \{ w=\sum_{A} z_A e_A , ~ z_A \in
    \mathbb{C}, A \subset N \right\},
    \end{eqnarray*}
where the imaginary unit interact with the basis elements as $i e_j
= e_j i,~j= 1, \ldots, n.$

Finally we introduce the conjugate of $w=\sum_{A} z_A e_A$ as
$\ov{w} = \sum_{A}
\ov{z_A}^{\mathbb{C}} \ov{e_A}^{\Cl_{0,n}}.$\\

Let now $\Omega \subset \BR^n \times \BR^+$ denote a bounded domain
with sufficiently smooth boundary $\Gamma=\partial\Omega.$ A
function $u: \Omega \rightarrow \Cl_n$ has a representation
$u=\sum_A u_A e_A$ with $\BC-$valued components $u_A.$ Properties
such as continuity will be understood component-wisely. In the
following we will use the short notation $L_p(\Omega),$
$C^k(\Omega),$ etc., instead of $L_p(\Omega,\Cl_n),$
$C^k(\Omega,\Cl_n).$ For more details see \cite{DSS}.

We consider the Dirac operator $D=\sum_{j=1}^{n} e_j\frac{\partial}
{\partial x_i}$ which has the property of factorizing the
$n$-dimensional Laplacian, that is, $D^2u=-\Delta u$. A
$\Cl_n$-valued function $u$ defined on an open domain $G,$ is said
to be {\it left-monogenic} if it satisfies $Du=0$ on $G.$

Taking into account \cite{CKS}, we will imbed $\mathbb{R}^n$ into
$\mathbb{R}^{n+2}$. For that purpose we add two new basis elements
$\f$ and $\fm$ satisfying
    \begin{eqnarray}
         {\f}^2 = {\fm}^2 = 0, &  \f \fm + \fm \f = 1, &
         \f e_j +e_j \f =  \fm e_j +e_j \fm = 0, j=1,\cdots,n.
         \label{regras}
    \end{eqnarray}
The set $\{ \f, \, \fm\}$ is said to be a Witt basis for
$\mathbb{R}^2$ and it will allows us to create a suitable
factorization of the Schr\"odinger operator where only partial
derivatives are used.


\subsection{Finite Difference Calculus}

For each $\emptyset \neq G \subset \BR^n$, $(n \geq 1)$, we denote
the associated discrete domain $G_h,$ for a fixed mesh size $h>0,$
as
    \begin{eqnarray*}
    G_h = \left\{ h\ul{m} = (hm_1,...,~hm_n) \in G:~ \ul{m} \in \BZ^n \right\}.
    \end{eqnarray*}

For a continuous function $f$ on $G$, we denote by $R_hf$ its
restriction to the lattice $G_h$ of mesh size $h > 0.$

In the following we consider functions restricted to the lattices
$\BR_h^3$, $\BR_\tau^+.$ We define the corresponding discrete
$l_1$-spaces in the usual way:
    \begin{eqnarray*}
    u \in l_1(\BR_h^3) & \Leftrightarrow & ||u||_{l_1(\BR_h^3)} =
    \sum_{\ul{m} \in \BZ^3} |u(h\ul{m})| h^3 ~~ < \infty \\
    & & \\
    v \in l_1(\BR_\tau^+) & \Leftrightarrow & ||v||_{l_1(\BR_\tau^+)} =
    \sum_{k \in \BN} |v(\tau k)| \tau ~~ < \infty.
    \end{eqnarray*}

For a discrete function $u : \BR^3_h \times \BR_\tau^+ \rightarrow
\BC^4 \sim \BC \otimes \BH,$ given as $u(h \ul m, k \tau )= (u^0,
u^1, u^2, u^3),$ we have the finite difference approximation for the
stationary Dirac operators given by {\small
\begin{eqnarray*}
    \begin{array}{ccc}
    D_h^{-+}u = \left(
    \begin{array}{c}
    -\partial_h^{-1}u^1-\partial_h^{-2}u^2-\partial_h^{-3}u^3\\
    \partial_h^{-1}u^0-\partial_h^{3}u^2+\partial_h^{2}u^3\\
    \partial_h^{-2}u^0+\partial_h^{3}u^1-\partial_h^{1}u^3\\
    \partial_h^{-3}u^0-\partial_h^{2}u^1+\partial_h^{1}u^2
    \end{array}
    \right),
    \end{array}
    \begin{array}{ccc} &
    D_h^{+-}u  =  \left(
    \begin{array}{c}
    -\partial_h^{1}u^1-\partial_h^{2}u^2-\partial_h^{3}u^3\\
    \partial_h^{1}u^0-\partial_h^{-3}u^2+\partial_h^{-2}u^3\\
    \partial_h^{2}u^0+\partial_h^{-3}u^1-\partial_h^{-1}u^3\\
    \partial_h^{3}u^0-\partial_h^{-2}u^1+\partial_h^{-1}u^2
    \end{array}
    \right),
    \end{array} \\ & \\
    \begin{array}{ccc}
u D_h^{-+} = \left(
    \begin{array}{c}
     -\partial_h^{-1}u^1-\partial_h^{-2}u^2-\partial_h^{-3}u^3\\
    \partial_h^{-1}u^0+\partial_h^{3}u^2-\partial_h^{2}u^3\\
    \partial_h^{-2}u^0-\partial_h^{3}u^1+\partial_h^{1}u^3\\
    \partial_h^{-3}u^0+\partial_h^{2}u^1-\partial_h^{1}u^2
    \end{array}
    \right),
    \end{array}
    \begin{array}{ccc} &
    u D_h^{+-}  =  \left(
    \begin{array}{c}
    -\partial_h^{1}u^1-\partial_h^{2}u^2-\partial_h^{3}u^3\\
    \partial_h^{1}u^0+\partial_h^{-3}u^2-\partial_h^{-2}u^3\\
    \partial_h^{2}u^0-\partial_h^{-3}u^1+\partial_h^{-1}u^3\\
    \partial_h^{3}u^0+\partial_h^{-2}u^1-\partial_h^{-1}u^2
    \end{array}
    \right),
    \end{array}
    \end{eqnarray*}}
where
    $$\partial_h^{\pm s}u^j = \frac{(u^j(h\ul{m}\pm
    h e_s,k\tau)-u^j(h\ul{m},k\tau))}{h},~j=0,1,2,3, ~s=1,2,3,$$
represent the spatial forward/backward difference operators. We
remark that these difference operators factorize the discrete
Laplacian, in the sense that
    \begin{eqnarray*}
    D_h^{+-}D_h^{-+} u & = D_h^{-+}D_h^{+-}u   = - \Delta_hu ~ I_4 & = \left( \sum_{s=1}^3
    \partial_h^{-s}\partial_h^s u^s \right) I_4,
    \end{eqnarray*}
where $I_4$ is the $4 \times 4$ identity matrix. In what follows we
denote $D_h^{+-}$ and $D_h^{-+}$ as forward/backward difference
Dirac operators.


We also have the following (forward) time difference operator (for
more details see \cite{GS} and \cite{GH})
    \begin{eqnarray*}
    \partial_{\tau}u^j (h\ul{m},k \tau ) & = &
    \frac{u^j(h\ul{m},
    \tau(k+1))-u^j(h\ul{m},\tau k)}{\tau},~j=0,\cdots,3.
    \end{eqnarray*}

Further, we use the notations
    \begin{eqnarray*}
    \delta_h(h\ul{m}) = \left\{ \begin{array}{ccc}
                        \frac{1}{h^3} & \mbox{if} & \ul{m}=(0,0,0) \\
                        0 & \mbox{if} & \ul{m} \neq (0,0,0)
                        \end{array}\right.
    ~~ \mbox{and} ~~
    \delta_\tau(\tau k) = \left\{ \begin{array}{ccc}
                        \frac{1}{\tau} & \mbox{if} & k=0 \\
                        0 & \mbox{if} & k \neq 0
                        \end{array}\right.
    \end{eqnarray*}
for the discrete Delta function.

We will consider the discrete Fourier transform introduced by
Stummel (see \cite{GS} for more details) with respect to $x,$
    \begin{eqnarray*}
    (\mathcal{F}_hu)(\xi,\cdot) & = & \left\{ \begin{array}{ccc}
                             \ds \frac{h^3}{(2\pi)^{\frac{3}{2}}} \sum_{h\ul{m} \in \BZ^3} u(h\ul{m},\cdot)
                             \exp(ih\ul{m}\xi) & \mbox{for}~ \xi \in Q_h \\
                             & & \\
                             \ds 0 & \mbox{otherwise}
                           \end{array} \right.,
    \end{eqnarray*}
where $\ds Q_h  =  \left\{ \xi=(\xi_1,\xi_2,\xi_3) \in
\BR^3:~-\frac{\pi}{h}<\xi_1,\xi_2,\xi_3<+\frac{\pi}{h} \right\}.$


\section{Explicit difference equation}

As it was indicated before, in this section we study the explicit
equation (\ref{*}). This will be done in two steps. In the first one
we study the existence of a discrete symbol of the operator in
(\ref{*}), which we use to construct $E_{h,\tau},$ a fundamental
solution for discrete explicit Schr\"odinger backward time-dependent
operator. In the second part we will estimate the norm
    \begin{eqnarray*}
    ||E_{h,\tau}-R_\tau R_h E||_{l_1(G_h\times\BR_\tau^+)},
    \end{eqnarray*}
which will allows to calculate its limit when $h$ and $\tau$ tend to
zero. Here, $E$ is a fundamental solution for the continuous
Schr\"odinger backward time-dependent operator, where $R_\tau R_h E$
its restriction to the lattice.


\subsection{Discrete Symbol of the Fundamental Solution}
Let us consider the equation (\ref{*}). In order to simplify the
resolution of this equation, we need to introduce the abbreviation
for the symbol of the discrete Laplace operator
    \begin{eqnarray*}
    d^2 & = & \frac{4}{h^2} \left( \sin^2\left( \frac{h\xi_1}{2}\right) + \sin^2\left( \frac{h\xi_2}{2}\right) +
    \sin^2\left( \frac{h\xi_3}{2}\right) \right).
    \end{eqnarray*}

Applying the discrete Fourier transform to (\ref{*}), we get the
equation
    \begin{eqnarray*}
    \left( \left( d^2\mathcal{F}_hE_{h,\tau} - i\partial_\tau \mathcal{F}_h E_{h,\tau} \right)\right)(\xi,t) & =
    & \frac{1}{(2\pi)^{\frac{3}{2}}}
    \delta_\tau(t) \chi_h(\xi),
    \end{eqnarray*}
with $\chi_\tau$ being the characteristic function of $Q_h,$ which
has the solution
    \begin{eqnarray}
    \left( \mathcal{F}_hE_{h,\tau}\right)(\xi,t)& = & \frac{i}{2\pi} H(t) \left(
    1 - i \tau d^2 \right)^{\frac{t}{\tau}-1} \chi_h(\xi).
    \label{2*}
    \end{eqnarray}

Using the restriction of the continuous inverse Fourier transform
$\mathcal{F}$, to the $\BR_h^3,$ $\mathcal{F}_h^{-1} =
R_h\mathcal{F},$ which acts as an inverse for $\mathcal{F}_h$, we
obtain
    \begin{eqnarray}
    E_{h,\tau}(h\ul{m},\tau k) & = & i H(\tau k) \left( \left( 1 - i \tau \Delta_h
    \right)^{k-1} \delta_h \right)(h\ul{m}), \label{3*}
    \end{eqnarray}
which is a fundamental solution of the discrete explicit
Schr\"odinger backward time-dependent equation.


\subsection{Convergence Result}
As it was indicated before, we will now estimate the norm
\begin{eqnarray*}
    ||E_{h,\tau}-R_\tau R_h E||_{l_1(G_h\times\BR_\tau^+)}.
    \end{eqnarray*}

This estimation will be done in two parts. In the first we will
consider the case of the time limited interval, i.e., we will
estimate our norm in $G_h \times ]0,T_0]_\tau,$ with $T_0 \in
\BR^+$. In the second part we will consider our norm in the $G_h
\times ]T_0,+\infty].$

\subsubsection{Case of the Limited Time Interval}
Initially we shall study now the behavior of
    \begin{eqnarray*}
    ||E_{h,\tau}-R_\tau R_h E||_{l_1(G_h\times]0,T_0]_\tau)}.
    \end{eqnarray*}

For this purpose we rewrite the equation (\ref{*}) in the form
    \begin{eqnarray}
    \begin{array}{l}
    \ds E_{h,\tau}(h\ul{m},\tau(k+1)) \\ \\
    \ds \qquad \quad = \left( 1 + \frac{6i\tau}{h^2}
    \right) E_{h,\tau}(h\ul{m},\tau k) \\ \\
    \ds \qquad \quad + \frac{i\tau}{h^2} \left[ E_{h,\tau}(h(m_1+1),hm_2,hm_3,\tau k)
    + E_{h,\tau}(h(m_1-1),hm_2,hm_3,\tau k)
    \right. \nonumber \\ \\
    \ds \qquad \quad + \left. E_{h,\tau}(hm_1,h(m_2+1),hm_3,\tau k) + E_{h,\tau}(hm_1,h(m_2-1),hm_3,\tau k)
    \right. \nonumber \\ \\
    \ds \qquad \quad + \left. E_{h,\tau}(hm_1,hm_2,h(m_3+1),\tau k) + E_{h,\tau}(hm_1,hm_2,h(m_3-1),\tau k)
    \right],
    \end{array}
    & \label{4*}
    \end{eqnarray}
which implies that $E_{h,\tau}$ is supported in a cone. From
(\ref{3*}) and (\ref{4*}) we get
    \begin{eqnarray*}
    ||E_{h,\tau}(\cdot,\tau)||_{l_1(\BR_h^3)} \leq 1 ~~~~
    ||E_{h,\tau}(\cdot,\tau k)||_{l_1(\BR_h^3)} \leq 1 ~~~~
    ||E_{h,\tau}(\cdot,\tau(k+1))||_{l_1(\BR_h^3)} \leq 1.
    \end{eqnarray*}

Further, let $T_0=\tau m_0$, with $m_0 \in \BN$ and $T_0 \in \BR^+$.
By addition with respect to $t,$ we get the estimation
    \begin{eqnarray}
    ||E_{h,\tau}||_{l_1(G_h \times ]0,T_0]_\tau)} ~~ \leq  ~~ ||E_{h,\tau}||_{l_1(\BR_h^3 \times
    ]0,T_0]_\tau)} ~~ \leq ~~ \sum_{k=1}^{m_0} \tau ~~ = ~~ T_0. \label{5*}
    \end{eqnarray}

Now we consider the continuous fundamental solution (see \cite{CV}
for more details)
    \begin{eqnarray}
    E(x,t) & = & \frac{iH(t)}{(4i\pi t)^{\frac{3}{2}}}~
    \exp\left(\frac{i|x|^2}{4t}\right) \label{solfuncon}
    \end{eqnarray}
of the continuous Schr\"odinger backward time-dependent operator.

We get
    \begin{eqnarray*}
    ||R_hE(\cdot,t)||_{l_1(G_h)} & \leq & H(t) \sum_{h\ul{m} \in
    G_h} \left| \frac{i}{(4i\pi t )^{\frac{3}{2}}}\right|~\left|
    \exp \left(\frac{ih^2|\ul{m}|^2}{4t}\right) \right|~h^3 \\
    & = & \frac{H(t)}{(4 \pi t)^{\frac{3}{2}}} \sum_{h\ul{m} \in
    G_h} h^3 \\
    & & \frac{H(t)}{(4 \pi t)^{\frac{3}{2}}}~Vol(G_h),
    \end{eqnarray*}
where $\ds Vol(G_h) = \sum_{h\ul{m} \in G_h} h^3.$ Furthermore
    \begin{eqnarray}
    ||R_\tau R_h E||_{l_1(G_h\times ]0,T_0]_\tau)} & =
    & \sum_{k=1}^{m_0} \frac{Vol(G_h)}{(4\pi \tau k)^{\frac{3}{2}}}~\tau \nonumber \\
    & = & \frac{Vol(G_h)}{(4\pi \tau)^{\frac{3}{2}}~\tau^{\frac{1}{2}}}~\sum_{k=1}^{m_0}
    \frac{1}{k^{\frac{3}{2}}} \nonumber \\
    & & \nonumber \\
    & \leq & \left\{ \begin{array}{ccc}
                       Vol(G_h)~\frac{T_0}{(4\pi)^{\frac{3}{2}}} & \mbox{if} & \tau \geq 1 \\
                        &  &  \\
                        Vol(G_h)~\frac{T_0+m_0^2}{(4\pi)^{\frac{3}{2}}} & \mbox{if} & 0 < \tau <1
                     \end{array}
    \right.. \label{6*}
    \end{eqnarray}

From (\ref{5*}) and (\ref{6*}) we conclude
    \begin{eqnarray}
    ||E_{h,\tau}-R_\tau R_h E||_{l_1(G_h \times ]0,T_0]_\tau)} & \leq &
    \left\{ \begin{array}{ccc}
              T_0 + Vol(G_h)~\frac{T_0}{(4\pi)^{\frac{3}{2}}} & \mbox{if} & \tau \geq 1 \\
              &  &  \\
              T_0 + Vol(G_h)~\frac{T_0+m_0^2}{(4\pi)^{\frac{3}{2}}} & \mbox{if} &
              0 < \tau < 1
            \end{array}
    \right.. \label{7*}
    \end{eqnarray}

The previous inequality describes the approximation error of the
fundamental solution (\ref{3*}) for small values of time variable.


\subsubsection{Case of the Unlimited Time Interval}

In the following, for $t=\tau k \in \BR_\tau^+$ with $t>T_0,$ we
study
    \begin{eqnarray*}
    ||E_{h,\tau}(\cdot,t)-R_\tau R_h E(\cdot,t)||_{l_1(G_h)}
    \end{eqnarray*}

In order to guarantee the convergence of some series and integrals,
we need to consider the following regularized fundamental solution
of the Schr\"odinger operator
    \begin{eqnarray*}
    E^\epsilon(x,t) & = & \frac{i H(t)}{(4i\pi
    t)^{\frac{3}{2}}}~\exp \left(\frac{(-\epsilon+i)|x|^2}{4t}\right),
    \end{eqnarray*}
which converges, in $L_p(G \times \BR^+),$ with $1 \leq p <
+\infty,$ to the continuous fundamental solution (\ref{solfuncon}).

We have
    \begin{eqnarray}
    \begin{array}{l}
    \ds ||E_{h,\tau}(\cdot,\tau k)-R_\tau R_h E(\cdot,\tau k)||_{l_1(G_h)} \\ \\
    \ds ~~~~ = \sum_{h\ul{m} \in G_h} \left| E_{h,\tau}(h\ul{m},\tau k)-R_\tau R_h E(h\ul{m},\tau k)
    \right|~h^3 \\ \\
    \ds ~~~~ \leq Vol(G_h)~\max_{h\ul{m} \in G_h} \left| E_{h,\tau}(h\ul{m},\tau k)-R_\tau R_h E(h\ul{m},\tau k)
    \right| \\ \\
    \ds ~~~~ = Vol(G_h)~\max_{h\ul{m} \in G_h} \left| (R_h\mathcal{F}\mathcal{F}_hE_{h,\tau})(h\ul{m},\tau k)
    - (R_\tau R_h \mathcal{F} \mathcal{F}^{-1} E)(h\ul{m},\tau k)
    \right| \\ \\
    \ds ~~~~ = Vol(G_h)~\max_{h\ul{m} \in G_h} \left[ \left| (R_h\mathcal{F}\mathcal{F}_hE_{h,\tau})(h\ul{m},\tau k)
    - (R_\tau R_h \mathcal{F} \mathcal{F}^{-1} E^\epsilon)(h\ul{m},\tau k)
    \right| \right. \\ \\
    \ds ~~~~ \qquad \qquad \qquad \qquad \qquad + \left. \left| (R_\tau R_h \mathcal{F}
    \mathcal{F}^{-1}E^\epsilon)(h\ul{m},\tau k) - (R_\tau R_h \mathcal{F} \mathcal{F}^{-1} E)(h\ul{m},\tau k)
    \right| \right] \\ \\
    \ds ~~~~ \leq Vol(G_h)~\left[\left| \frac{1}{2\pi} \int_{\BR^3}\left[(\mathcal{F}_hE_{h,\tau})(x,\tau k)
    -(R_\tau \mathcal{F}^{-1}
    E)(x,\tau k) \right]~\exp \left(-ix\xi\right) d\xi \right| \right. \\ \\
    \ds ~~~~ \qquad \qquad \qquad \qquad \qquad + \left. \max_{h\ul{m} \in G_h} \left| (R_\tau R_h \mathcal{F}
    \mathcal{F}^{-1}E^\epsilon)(h\ul{m},\tau k)-(R_\tau R_h \mathcal{F} \mathcal{F}^{-1} E)(h\ul{m},\tau k)
    \right| \right] \\ \\
    \ds ~~~~ \leq  \frac{Vol(G_h)}{2\pi}~\left[\int_{\BR^3}\left|(\mathcal{F}_hE_{h,\tau})(x,\tau k)
    -(R_\tau \mathcal{F}^{-1}
    E)(x,\tau k) \right|~\left|\exp \left(-ix\xi\right) \right| d\xi \right. \\ \\
    \ds ~~~~ \qquad \qquad \qquad \qquad \qquad + \left. \max_{h\ul{m} \in G_h}~  \left| (R_\tau R_h
    \mathcal{F} \mathcal{F}^{-1}E^\epsilon)(h\ul{m},\tau k)-(R_\tau R_h \mathcal{F} \mathcal{F}^{-1} E)(h\ul{m},\tau k)
    \right| \right] \\ \\
    \ds ~~~~ = \frac{Vol(G_h)}{2\pi}~ \left[\underbrace{||(\mathcal{F}_hE_{h,\tau})(\cdot,\tau k)-(R_\tau
    \mathcal{F}^{-1}E^\epsilon)(\cdot,\tau k)||_{L_1(\BR^3)}}_{(I)}
    \right.\\ \\
    \ds ~~~~ \qquad \qquad \qquad \qquad \qquad \left. +
    \underbrace{\max_{h\ul{m} \in G_h}\left| (R_\tau R_h \mathcal{F} \mathcal{F}^{-1}E^\epsilon)(h\ul{m},\tau k)
    -(R_\tau R_h \mathcal{F} \mathcal{F}^{-1} E)(h\ul{m},\tau k)
    \right|}_{(II)} \right].
    \end{array} &
    \label{8*}
    \end{eqnarray}

By the convergence of the regularized fundamental solution
$E^\epsilon$ to the continuous one $E$, we conclude immediately that
the term ($II$) converges to zero, as $\epsilon$ goes to $0^+$. This
fact implies that the study of (\ref{8*}) depends on the analysis of
the term ($I$). In this case, we need to split this study into the
outside of $Q_h$ and its inside.

In $\BR^3 \setminus Q_h$, we have for the term ($I$)
    \begin{eqnarray}
    \begin{array}{l}
    \ds ||(\mathcal{F}_hE_{h,\tau})(\cdot,\tau k)-(R_\tau
    \mathcal{F}^{-1}E^\epsilon)(\cdot,\tau k)||_{L_1(\BR^3)} \\ \\
    \ds ~~~~ = ||(R_\tau \mathcal{F}^{-1}E^\epsilon)(\cdot,\tau k)||_{L_1(\BR^3)} \\ \\
    \ds ~~~~ = \frac{H(\tau k)}{(4\pi t)^{\frac{3}{2}}}~
    \frac{(2\tau k)^3}{\epsilon^{\frac{3}{2}}}~\left| \left|
    \exp \left(-\frac{\tau k|\xi|^2}{\epsilon}\right)~\exp \left(-i\left( \tau k|\xi|^2 - \frac{\pi}{4}
    \right)\right)\right| \right|_{L_1(\BR^3 \setminus Q_h)} \\ \\
    \ds ~~~~ = \frac{H(\tau k)~2~\epsilon^{\frac{3}{2}}}{(\pi
    \tau k)^{\frac{3}{2}}}~\exp \left(-\frac{\tau k\pi^2}{h^2\epsilon}\right).
    \end{array} & \label{9*}
    \end{eqnarray}

In $Q_h$ we have for ($I$) the estimation
    \begin{eqnarray*}
    \begin{array}{l}
    \ds ||(\mathcal{F}_hE_{h,\tau})(\cdot,\tau k)-(R_\tau
    \mathcal{F}^{-1}E^\epsilon)(\cdot,\tau k)||_{L_1(\BR^3)} \\ \\
    \ds ~~~~ = \frac{H(\tau k)}{(2\pi)^{\frac{3}{2}}} \left| \left| (1-i\tau d^2)^k -
    \left( \frac{2\tau k}{\epsilon} \right)^{\frac{3}{2}}
    \exp\left( -\frac{\tau k|\xi|^2}{\epsilon}\right)
    \exp\left( -i \left( \tau k|\xi|^2-\frac{3\pi}{4}\right)\right)\right|
    \right|_{L_1(\BR^3)} \\ \\
    \ds ~~~~ = \frac{H(\tau k)}{(2\pi)^{\frac{3}{2}}} \left( \frac{2\tau k}{\epsilon} \right)^{\frac{3}{2}} \\ \\
    \ds ~~~~ ~~~~ \cdot \left| \left|
    \left( \frac{\epsilon}{2\tau k} \right)^{\frac{3}{2}}~(1-i\tau d^2)^k
    - \exp\left( -\frac{\tau k|\xi|^2}{\epsilon}\right)
    \exp\left( -i \left( \tau k|\xi|^2-\frac{3\pi}{4}\right)\right)\right|
    \right|_{L_1(\BR^3)}
    \end{array}
    \end{eqnarray*}
    \begin{eqnarray}
    \begin{array}{l}
    \ds ~~~~ \leq \frac{H(\tau k)}{(2\pi)^{\frac{3}{2}}} ~ \left(\frac{2\tau k}{\epsilon} \right)^{\frac{3}{2}} \\ \\
    \ds ~~~~ ~~~~ \cdot \left[ \underbrace{\left| \left|
    \left( \frac{\epsilon}{2\tau k} \right)^{\frac{3}{2}}~
    (1-i\tau d^2)^k - \exp\left( -\frac{\tau kd^2}{\epsilon}\right)
    \exp\left( -i \left( \tau kd^2-\frac{3\pi}{4}\right)\right)\right|
    \right|_{L_1(\BR^3)}}_{(III)}\right. \\ \\
    \ds \qquad \qquad \qquad  + \left| \left|
    \exp\left( -\frac{\tau kd^2}{\epsilon}\right)^{\frac{3}{2}}
    \exp\left( -i \left( \tau kd^2-\frac{3\pi}{4}\right)\right) \right. \right. \\ \\
    \ds \qquad \qquad \qquad \qquad  \underbrace{\left.  \left. \left.
    - \exp\left( -\frac{\tau k|\xi|^2}{\epsilon}\right)^{\frac{3}{2}}
    \exp\left( -i \left( \tau k|\xi|^2-\frac{3\pi}{4}\right)\right)\right|
    \right|_{L_1(\BR^3)}\right]}_{(IV)}
    \end{array} & \label{10*}
    \end{eqnarray}

From
    \begin{eqnarray*}
    \frac{|\xi|^2-d^2}{\epsilon} & = & \left[ |\xi|^2 - \frac{4}{h^2} \left( \sin^2\left( \frac{h\xi_1}{2}\right) +
    \sin^2\left( \frac{h\xi_2}{2}\right)+\sin^2\left( \frac{h\xi_3}{2}\right)\right)
    \right] ~ \frac{1}{\epsilon} \\
    & \leq & \left[ \frac{h^2}{12} ( \xi_1^4 + \xi_2^4 + \xi_3^4 ) \right] ~
    \frac{1}{\epsilon} \\
    & \leq & \frac{h^2|\xi|^2}{12\epsilon},
    \end{eqnarray*}
we get
    $$\begin{array}{l}
    \ds \left| \exp\left( -\frac{\tau kd^2}{\epsilon}\right)^{\frac{3}{2}}
    \exp\left( -i \left( \tau kd^2-\frac{3\pi}{4}\right)\right)
    - \exp\left( -\frac{\tau k|\xi|^2}{\epsilon}\right)^{\frac{3}{2}}
    \exp\left( -i \left( \tau k|\xi|^2-\frac{3\pi}{4}\right)\right)
    \right| \\ \\
    \ds ~~~~ \leq \left| \exp\left( -\frac{\tau kd^2}{\epsilon}\right) - \exp\left(
    -\frac{\tau k|\xi|^2}{\epsilon}\right)\right|~
    \left| \exp\left( -i \left( \tau k|\xi|^2 - \frac{3\pi}{4} \right) \right)
    \right| \\ \\
    \ds \qquad \qquad + \left| \exp\left( -\frac{\tau k|\xi|^2}{\epsilon}\right)
    \right| ~ \left| \exp\left( -i \left( \tau kd^2 - \frac{3\pi}{4} \right) \right) -
    \exp\left( -i \left( \tau k|\xi|^2 - \frac{3\pi}{4} \right) \right)
    \right| \\ \\
    \ds ~~~~ \leq  \left| \exp\left( -\frac{\tau kd^2}{\epsilon}\right) - \exp\left(
    -\frac{\tau k|\xi|^2}{\epsilon}\right)\right| + 2
    \exp\left(-\frac{\tau k|\xi|^2}{\epsilon}\right) \\ \\
    \ds ~~~~ \leq \tau k \left( \frac{|\xi|^2-d^2}{\epsilon}\right)~\exp\left(
    - \frac{\tau kd^2}{\epsilon} \right) + 2
    \exp\left(-\frac{\tau k|\xi|^2}{\epsilon}\right) \\ \\
    \ds ~~~~ \leq \frac{\tau kh^2|\xi|^2}{12\epsilon}~\exp\left( -\frac{4|\xi|^2\tau k}{\pi^2\epsilon}
    \right) + 2
    \exp\left(-\frac{\tau k|\xi|^2}{\epsilon}\right),
    \end{array}$$
which implies that ($IV$) satisfies
    \begin{eqnarray}
    (IV) & \leq & \frac{\tau kh^2}{3\epsilon} \int_0^{\frac{\sqrt2
    \pi}{h}} \int_0^{\frac{\pi}{2}}
    \int_{-\frac{\pi}{h}}^{\frac{\pi}{h}}
    r^5~\exp\left( -\frac{4r^2\tau k}{\pi^2\epsilon} \right) dz d\varphi
    dr \nonumber \\
    & & ~~~~ ~~~~ + 2 \int_0^{\frac{\sqrt2
    \pi}{h}} \int_0^{\frac{\pi}{2}} \int_{-\frac{\pi}{h}}^{\frac{\pi}{h}}
    \exp\left( -\frac{(\tau k)^2}{\epsilon} \right) dz d\varphi
    dr \nonumber \\
    & \leq & \frac{\pi^8}{192} \left[ \exp\left(-\frac{8\tau k}{\epsilon h^2}\right)
    \left( \frac{\epsilon^2h}{(\tau k)^2} + \frac{8\epsilon}{\tau kh} +
    \frac{32}{h^3}\right) - \frac{\epsilon^2h}{(\tau k)^2} \right]
    \nonumber \\
    & & ~~~~ ~~~~ + \frac{2\pi^2}{h}~ \left( \frac{\pi\epsilon}{\tau k(h^2-1)}\right)^{\frac{1}{2}}~
    \int_0^{2\sqrt{\frac{2\tau k}{\epsilon h}}} \exp(-y^2) dy.  \label{aux1}
    \end{eqnarray}

Also, we have the following relation
    $$\begin{array}{l}
    \ds \left| \left( \frac{\epsilon}{2\tau k} \right)^{\frac{3}{2}}~
    (1-i\tau d^2)^k - \exp\left( -\frac{\tau kd^2}{\epsilon}\right)
    \exp\left( -i \left( \tau kd^2-\frac{3\pi}{4}\right)\right)\right| \\ \\
    \ds ~~~~ \leq \left( \frac{\epsilon}{2\tau k} \right)^{\frac{3}{2}}~
    |1-i\tau d^2|^k + \exp\left(
    -\frac{\tau kd^2}{\epsilon}\right) \\ \\
    \ds ~~~~ \leq \left( \frac{\epsilon}{2\tau k} \right)^{\frac{3}{2}}~
    (1+\tau^2 d^4)^k +
    \exp\left(-\frac{\tau kd^2}{\epsilon}\right) \\ \\
    \ds ~~~~ \leq \left( \frac{\epsilon}{2\tau k} \right)^{\frac{3}{2}}~
    (1+\tau^2(h^2-1)^2|\xi|^4)^k +
    \exp\left(-\frac{\tau k(h^2-1)|\xi|^2}{\epsilon}\right),
    \end{array}$$
which implies that
    \begin{eqnarray}
    (III) & \leq & \left( \frac{\epsilon}{2\tau k} \right)^{\frac{3}{2}}
    \int_0^{\frac{\sqrt2 \pi}{h}} \int_0^{\frac{\pi}{2}}
    \int_{-\frac{\pi}{h}}^{\frac{\pi}{h}}
    (1+\tau^2(h^2-1)^2r^4)^k  dz d\varphi
    dr \nonumber \\
    & & ~~~~ ~~~~ + \int_0^{\frac{\sqrt2
    \pi}{h}} \int_0^{\frac{\pi}{2}} \int_{-\frac{\pi}{h}}^{\frac{\pi}{h}}
    \exp\left(-\frac{\tau k(h^2-1)r^2}{\epsilon}\right) dz d\varphi
    dr \nonumber \\
    & = & \frac{\epsilon \pi^2}{\tau kh} \int_0^{\frac{\sqrt2 \pi}{h}}
    (1+\tau^2(h^2-1)^2r^4)^k dr \nonumber \\
    & & ~~~~ ~~~~ + \frac{\pi^2}{h} \int_0^{\frac{\sqrt2
    \pi}{h}} \exp\left(-\frac{\tau k(h^2-1)r^2}{\epsilon}\right) dr \nonumber \\
    & = & \frac{\epsilon \pi^2}{\tau kh}~C_1(\epsilon,h,\tau) +
    \frac{\pi^2}{2h}~\left( \frac{\pi\epsilon}{\tau k(h^2-1)}
    \right)^{\frac{1}{2}}~\int_0^{\frac{\pi}{h}~\sqrt{\frac{2\tau k(h^2-1)}{\epsilon}}}
    \exp(-y^2) dy, \nonumber \\
    & & \label{aux2}
    \end{eqnarray}
where
    \begin{eqnarray*}
    C_1(\epsilon,h,\tau) & = & \int_0^{\frac{\sqrt2 \pi}{h}}
    (1+\tau^2(h^2-1)^2r^4)^k dr
    \end{eqnarray*}
tends to zero when $\epsilon < \tau^8$ and $\ds
\frac{\tau}{h^2}<\frac{1}{6\pi^2}$.

Taking into account the estimates obtained previously for the terms
($III$) and ($IV$) we conclude that inside $Q_h$ we have
    \begin{eqnarray}
    (\ref{10*}) & \leq & \frac{H(t)}{(2\pi)^{\frac{3}{2}}} ~ \left(\frac{2\tau k}{\epsilon}
    \right)^{\frac{3}{2}} \left[ \frac{\epsilon \pi^2}{\tau kh}~C_1(\epsilon,h,\tau) \right. \nonumber \\
    & & \qquad \qquad \qquad \left. + \frac{\pi^2}{2h}~\sqrt{\frac{\pi\epsilon}{\tau k(h^2-1)}}~
    \int_0^{\frac{\pi}{h}~\sqrt{\frac{2\tau k(h^2-1)}{\epsilon}}}
    \exp(-y^2) dy \right. \nonumber \\
    & & \qquad \qquad \qquad + \left. \frac{\pi^8}{192} \left[ \exp\left(-\frac{8\tau k}{\epsilon h^2}\right)
    \left( \frac{\epsilon^2h}{(\tau k)^2} + \frac{8\epsilon}{\tau kh} +
    \frac{32}{h^3}\right) - \frac{\epsilon^2h}{(\tau k)^2} \right] \right.
    \nonumber \nonumber \\
    & & \qquad \qquad \qquad \left. + \frac{2\pi^2}{h}~\sqrt{
    \frac{\epsilon}{\tau k}}
    \int_0^{\frac{2}{h}\sqrt{\frac{2\tau k}{\epsilon}}}
    \exp(y^2) dy,\right]. \label{aux3}
    \end{eqnarray}

Taking into account the estimations obtained in (\ref{9*}) and
(\ref{aux3}), we conclude that
    \begin{eqnarray}
    \begin{array}{l}
    \ds ||E_{h,\tau}(\cdot,\tau k)-R_\tau R_h E(\cdot,\tau k)||_{l_1(G_h)} \\ \\
    \ds ~~~~ = \frac{Vol(G_h)}{2\pi}~ \left[ 2H(\tau k)~\left(\frac{\epsilon}{\pi
    \tau k}\right)^{\frac{3}{2}}~\exp \left(-\frac{\tau k\pi^2}{h^2\epsilon}\right)
    \right.\\ \\
    \ds \qquad  \qquad \qquad \qquad   +  H(\tau k)~\left(\frac{2\tau k}{2\pi \epsilon}
    \right)^{\frac{3}{2}} \left[ \frac{\epsilon \pi^2}{\tau kh}~C_1(\epsilon,h,\tau) \right.
    \\ \\
    \ds \qquad \qquad \qquad \qquad \qquad \left. + \frac{\pi^2}{2h}~\sqrt{\frac{\pi\epsilon}{\tau k(h^2-1)}}
    ~\int_0^{\frac{\pi}{h}~\sqrt{\frac{2\tau k(h^2-1)}{\epsilon}}}
    \exp(-y^2) dy \right. \\ \\
    \ds \qquad \qquad \qquad \qquad \qquad
    + \left. \frac{\pi^8}{192} \left[ \exp\left(-\frac{8\tau k}{\epsilon h^2}\right)
    \left( \frac{\epsilon^2h}{(\tau k)^2} + \frac{8\epsilon}{\tau kh} +
    \frac{32}{h^3}\right) - \frac{\epsilon^2h}{(\tau k)^2} \right] \right.
    \\ \\
    \ds \qquad \qquad  \qquad \qquad \qquad \left.\left. + \frac{2\pi^2}{h}~
    \sqrt{\frac{\epsilon}{\tau k}}
    \int_0^{\frac{2}{h}~\sqrt{\frac{2\tau k}{\epsilon}}}
    \exp(y^2) dy,\right]\right].
    \end{array} &
    \label{aux4}
    \end{eqnarray}

Evaluating now the $l_1-$norm with respect to the time-lattice
    \begin{eqnarray*}
    \begin{array}{l}
    \ds ||E_{h,\tau}(\cdot,\tau k)-R_\tau R_h E(\cdot,\tau k)||_{l_1(G_h \times (T_0,+\infty)_\tau)} \\ \\
    \ds ~~~~ = \left| \left|~||E_{h,\tau}(\cdot,\tau k)-R_\tau R_h E(\cdot,\tau
    k)||_{l_1(G_h)} \right| \right|_{l_1((T_0,+\infty)_\tau)} \\ \\
    \ds ~~~~ = \sum_{k=m_0+1}^{+\infty} \tau ||E_{h,\tau}(\cdot,\tau k)-R_\tau R_h E(\cdot,\tau
    k)||_{l_1(G_h)}
    \end{array}
    \end{eqnarray*}
    \begin{eqnarray}
    \begin{array}{l}
    \ds ~~~~ = \frac{Vol(G_h)}{2\pi}~ \sum_{k=m_0+1}^{+\infty} \left[ 2~\left(\frac{2~\epsilon}{\pi
    k\tau}\right)^{\frac{3}{2}}~\exp \left(-\frac{k\tau\pi^2}{h^2\epsilon}\right)
    +  \left(\frac{k\tau}{\pi\epsilon}
    \right)^{\frac{3}{2}} \left[ \frac{\epsilon \pi^2}{k\tau h}~C_1(\epsilon,h,\tau) \right. \right.\\ \\
    \ds \qquad \qquad \qquad \qquad \qquad \qquad \left. + \frac{\pi^2}{2h}~\sqrt{\frac{\pi\epsilon}{k\tau(h^2-1)}}
    ~\int_0^{\frac{\pi}{h}~\sqrt{\frac{2k\tau(h^2-1)}{\epsilon}}}
    \exp(-y^2) dy \right. \\ \\
    \ds \qquad \qquad \qquad \qquad \qquad \qquad
    + \left. \frac{\pi^8}{192} \left[ \exp\left(-\frac{8k\tau}{\epsilon h^2}\right)
    \left( \frac{\epsilon^2h}{(k\tau)^2} + \frac{8\epsilon}{k\tau h} +
    \frac{32}{h^3}\right) - \frac{\epsilon^2h}{(k\tau)^2} \right] \right.
    \\ \\
    \ds \qquad \qquad  \qquad \qquad \qquad \qquad \left.\left. + \frac{2\pi^2}{h}~
    \sqrt{\frac{\epsilon}{k\tau}}~
    \int_0^{\frac{2}{h}~\sqrt{\frac{2\epsilon}{k\tau}}}
    \exp(-y^2) dy\right]\right].
    \end{array} &
    \label{17*}
    \end{eqnarray}

After straightforward  calculations we conclude that the previous
series is convergent and its sums, which we will denote by
$C_2(h,h,\tau)$, tends to zero when $\epsilon<\tau^8$ and $\ds
\frac{\tau}{h^2}<\frac{1}{6\pi^2}$.


\subsubsection{Main Result}

Using the inequalities (\ref{7*}) and (\ref{17*}) we obtain the
general estimation
    \begin{eqnarray*}
    & ||E_{h,\tau}-R_\tau R_h E||_{l_1(G_h \times
    \BR_\tau^+)} \\ & \\
    \leq &
    \left\{ \begin{array}{ccc}
              T_0 + Vol(G_h)~\frac{T_0}{(4\pi)^{\frac{3}{2}}} + C_2(\epsilon,h,\tau) & \mbox{if} & \tau \geq 1 \wedge
              \left( \epsilon<\tau^8 \wedge \frac{\tau}{h^2}<\frac{1}{6\pi^2} \right) \\
              &  &  \\
              T_0 + Vol(G_h)~\frac{T_0+m_0^2}{(4\pi)^{\frac{3}{2}}} + C_2(\epsilon,h,\tau) & \mbox{if} &
              0 < \tau < 1 \wedge \left( \epsilon<\tau^8 \wedge
              \frac{\tau}{h^2}<\frac{1}{6\pi^2} \right)
            \end{array}
    \right.. \label{18*}
    \end{eqnarray*}

For the purpose of our convergence theorem we require that $h \leq
h_0$, where $h_0$ is an arbitrary constant. Now we can formulate the
following convergence theorem
    \begin{theo}\label{convteo1}
    Let $\ds \frac{\tau^2}{h^2} < \frac{1}{6\pi^2}$. Then
        \begin{eqnarray*}
        \left| \left| E_{h,\tau} - R_\tau R_h E \right| \right|_{l_1(G_h \times \BR_\tau^+)}
        \rightarrow 0 \quad \mbox{for}~ h,\tau \rightarrow 0^+.
        \end{eqnarray*}
    \end{theo}
    \begin{proof}
    We prove that for an arbitrary $\delta >0$ there exists a
    constant $h$ such that
        \begin{eqnarray*}
        \left| \left| E_{h,\tau} - R_\tau R_h E \right| \right|_{l_1(G_h \times
        \BR_\tau^+)} & \leq & \delta
        \end{eqnarray*}
    Also, it is possible to choose a $T_0$ in the
    lattice $\BR_\tau^+$, and define
        \begin{eqnarray*}
        T_0^+ = T_0+\alpha\tau \qquad \mbox{and} \qquad
        T_0^-=T_0-(1-\alpha)\tau \qquad \mbox{with } \alpha \in
        [0,1)
        \end{eqnarray*}
    such that $T_0^+ \in \BR_\tau^+$ and $T_0^- \in \BR_\tau^+$. Obviously
    we have
        $$\begin{array}{l}
        \ds \left| \left| E_{h,\tau} - R_\tau R_h E \right| \right|_{l_1(G_h \times
        \BR_\tau^+)} \\ \\
        \qquad \leq \left| \left| E_{h,\tau} - R_\tau R_h E \right| \right|_{l_1(G_h \times
        ]0,T_0^+]_\tau)} + \left| \left| E_{h,\tau} - R_\tau R_h E \right| \right|_{l_1(G_h \times
        (T_0^-,+\infty)_\tau)}.
        \end{array}$$
    Now, a simple estimation using (\ref{7*}) and (\ref{17*}) shows
    that the right-hand side of the last inequality is bounded by
    $\delta$.
    \end{proof}



\section{Implicit Difference Equation}

In this section we will make a similar study for the implicit
equation, i.e, we will initially obtain the fundamental solution
$E_{h,\tau}^\ast$ of the following implicit equation
    \begin{eqnarray}
    (-\Delta_hE_{h,\tau}^\ast)(h\ul{m},\tau(k+1)) - i (\partial_\tau E_{h,\tau}^\ast)(h\ul{m},\tau k) & = &
    \delta_h(h\ul{m})~\delta_\tau(\tau k), \label{aux5}
    \end{eqnarray}
and then we will develop similar convergent results for this
solution.

Using the discrete Fourier transform again we find the solution
    \begin{eqnarray*}
    (\mathcal{F}_hE_{h,\tau}^\ast)(\xi,t) & = & \frac{i}{2\pi}~\left(1+i\tau
    d^2\right)^{-\frac{t}{\tau}} \chi_h(\xi)
    \end{eqnarray*}
in analogy to (\ref{2*}). Finally, we get the following system of
equations to calculate $E_{h,\tau}^\ast(x,t)$
    \begin{eqnarray}
    \left\{
    \begin{array}{rclcc}
      \ds E_{h,\tau}^\ast(h\ul{m},0)& = & 0 & \mbox{for} & h\ul{m} \in \BR_h^3 \\
      & & & & \\
      \ds ((1-i\tau\Delta_h)E_{h,\tau}^\ast)(h\ul{m},\tau k) & =
      & \delta_h(h\ul{m}) & \mbox{if} & h\ul{m} \in \BR_h^3 \\
      & & & & \\
      \ds ((1-i\tau\Delta_h)E_{h,\tau}^\ast)(h\ul{m},\tau(k+1)) & =
      & E_{h,\tau}^\ast(h\ul{m},\tau k) & \mbox{if} & h\ul{m} \in
      \BR_h^3,~\tau k\in\BR_0^+,~k \in \BZ^+
    \end{array}
    \right..
    & \label{19*}
    \end{eqnarray}

We note that it is also possible to describe the fundamental
solution by application of $R_hF$
    \begin{eqnarray*}
    E_{h,\tau}^\ast(h\ul{m},\tau k) & = & R_h\mathcal{F} \left( \frac{H(t)}{2\pi} (1+i\tau d^2)^{-\frac{t}{\tau}}
    \chi_h(\xi)\right)(h\ul{m},\tau k).
    \end{eqnarray*}

However, this does not proves our assertion, that of
$E_{h,\tau}^\ast$ being a fundamental solution of (\ref{*}). For
that purpose, we need to do a similar study as in Section 2.2.


\subsection{Existence of Fundamental Solution}

First we have the following three lemmas
    \begin{lem}\label{lema1}
    Let $f_h$ be an arbitrary bounded function. Then the equation
        \begin{eqnarray*}
        (1-i\tau\Delta_h) v_h & = & f_h(x),
        \end{eqnarray*}
    for all $x \in \BR_h^3$ has a unique solution $v_h$.
    \end{lem}
    \begin{proof}
    We will omit the presentation of this proof because it is very similar to the proof of the
    Lemma 1 presented in \cite{GH}.
    \end{proof}

The following two lemmas are proved in \cite{GH}.
    \begin{lem}\label{lema2}
    If $f_h \in l_1(\BR^3_h)$, then $v_h \in l_1(\BR_h^3)$.
    \end{lem}

    \begin{lem}\label{lema3}
    If $|f_h(x)|<K_1e^{-c_1|x|}$ and $|e_h(x)|\leq K_2 e^{-c_2|x|}$
    with $0 < c_1 < c_2$, then $v_h(x) \leq K_5
    e^{-(c_1-\delta)|x|}$ for all $\delta > 0$.
    \end{lem}

With this three results we can present the following result about
the existence of the fundamental solution.
    \begin{theo}
    System (\ref{19*}) has a unique solution $E_{h,\tau}^\ast$ and for
    arbitrary $T_0<\infty$ it holds $E_{h,\tau}^\ast \in l_1(\BR_h^3 \times
    [0,T_0])$.
    \end{theo}
    \begin{proof}
    The assertion follows from Lemmas \ref{lema1}, \ref{lema2} and \ref{lema3}. We remark that the
    consideration in Lemma \ref{lema3} can be repeated as long as necessary.
    An estimation of the $l_1-$norm with respect to $t$ is possible
    because the number of time steps is bounded.
    \end{proof}


\subsection{Convergence}

For $t=0$ we can write the difference equation in the form
    $$\begin{array}{l}
    \ds \left( -i + \frac{6\tau}{h^2}\right) E_{h,\tau}^\ast (h\ul{m},\tau k)  =
    \delta_h(h\ul{m}) \\ \\
    \ds \qquad \qquad + \frac{\tau}{h^2} \left[ E_{h,\tau}^\ast(h(m_1+1),hm_2,hm_3,\tau k)
    + E_{h,\tau}^\ast(h(m_1-1),hm_2,hm_3,\tau k)
    \right. \\ \\
    \ds \qquad \qquad \left. + E_{h,\tau}^\ast(hm_1,h(m_2+1),hm_3,\tau k) + E_{h,\tau}^\ast(hm_1,h(m_2-1),hm_3,\tau k)
    \right. \\ \\
    \ds \qquad \qquad \left. + E_{h,\tau}^\ast(hm_1,hm_2,h(m_3+1),\tau k)
    + E_{h,\tau}^\ast(hm_1,hm_2,h(m_3-1),\tau k) \right],
    \end{array}$$

We have that
    \begin{eqnarray*}
    \left| -i + \frac{6\tau}{h^2}\right| \sum_{\ul{m} \in \BR_h^3}
    \left|E_{h,\tau}^\ast (h\ul{m},\tau) \right| h^3 & \leq &
    1 + \frac{6\tau}{h^2} \sum_{\ul{m} \in \BR_h^3} \left| E_{h,\tau}^\ast (h\ul{m},\tau) \right| h^3,
    \end{eqnarray*}
which implies that
    \begin{eqnarray*}
    ||E_{h,\tau}^\ast(\cdot,\tau)||_{l_1(\BR_h^3)} & \leq & 1.
    \end{eqnarray*}

In the same way we prove that the inequality
    \begin{eqnarray*}
    ||E_{h,\tau}^\ast(\cdot,\tau(k+1))||_{l_1(\BR_h^3)} & \leq &
    ||E_{h,\tau}^\ast(\cdot,\tau k)||_{l_1(\BR_h^3)},
    \end{eqnarray*}
for each $k \geq 1$ starting with the corresponding equations
(\ref{19*}). We obtain
    \begin{eqnarray*}
    ||E_{h,\tau}^\ast||_{l_1(G_h \times ]0,T_0]_\tau)} & \leq & ||E_{h,\tau}^\ast||_{l_1(\BR_h^3 \times
    ]0,T_0]_\tau)} \\
    & \leq & \sum_{k=1}^{m_0} \tau ~ = ~T_0.
    \end{eqnarray*}

In relation to the continuous fundamental solution we have the
result (\ref{6*})
    \begin{eqnarray}
    ||R_\tau R_h E||_{l_1(G_h \times ]0,T_0]_\tau)} & =
    & \sum_{k=1}^{m_0} \frac{Vol(G_h)}{(4\pi k\tau)^{\frac{3}{2}}}~\tau \nonumber \\
    & = & \frac{Vol(G_h)}{(4\pi \tau)^{\frac{3}{2}}~\tau^{\frac{1}{2}}}~\sum_{k=1}^{m_0}
    \frac{1}{k^{\frac{3}{2}}} \nonumber \\
    & & \nonumber \\
    & \leq & \left\{ \begin{array}{ccc}
                       Vol(G_h)~\frac{T_0}{(4\pi)^{\frac{3}{2}}} & \mbox{if} & \tau \geq 1 \\
                        &  &  \\
                        Vol(G_h)~\frac{T_0+m_0^2}{(4\pi)^{\frac{3}{2}}} & \mbox{if} & 0 < \tau <1
                     \end{array}
    \right.. \label{22*}
    \end{eqnarray}

Now we study
    \begin{eqnarray*}
    ||E_{h,\tau}^\ast(\cdot,\tau k) - R_h E (\cdot,\tau k)||_{l_1(G_h)},
    \end{eqnarray*}
para $k>m_0$. In order to estimate the right hand side of (\ref{8*})
we need to study the following inequality, which is very similar to
(\ref{10*})
\begin{eqnarray}
    \begin{array}{l}
    \ds ||(\mathcal{F}_hE_{h,\tau}^\ast)(\cdot,\tau k)-(R_\tau R_h
    \mathcal{F}^{-1}E^\epsilon)(\cdot,\tau k)||_{L_1(\BR^3)} \\ \\
    \ds ~~~~ \leq \frac{H(\tau k)}{(2\pi)^{\frac{3}{2}}} ~ \left(\frac{2\tau k}{\epsilon} \right)^{\frac{3}{2}} \\ \\
    \ds \qquad \cdot \left[ \underbrace{\left| \left|
    \left( \frac{\epsilon}{2\tau k} \right)^{\frac{3}{2}}~
    (1+i\tau d^2)^{-k} - \exp\left( -\frac{\tau kd^2}{\epsilon}\right)
    \exp\left( -i \left( \tau kd^2-\frac{3\pi}{4}\right)\right)\right|
    \right|_{L_1(\BR^3)}}_{(V)}\right. \\ \\
    \ds \qquad \qquad  + \left| \left|
    \exp\left( -\frac{\tau kd^2}{\epsilon}\right)^{\frac{3}{2}}
    \exp\left( -i \left( \tau kd^2-\frac{3\pi}{4}\right)\right) \right. \right. \\ \\
    \ds \qquad \qquad \qquad
    \underbrace{\left.  \left. \left. - \exp\left( -\frac{\tau k|\xi|^2}{\epsilon}\right)^{\frac{3}{2}}
    \exp\left( -i \left( \tau k|\xi|^2-\frac{3\pi}{4}\right)\right)\right|
    \right|_{L_1(\BR^3)}\right]}_{(VI)}.
    \end{array} & \label{23*}
    \end{eqnarray}

Taking into account that $(VI)=(IV)$ and the following relation
    $$\begin{array}{l}
    \ds \left| \left( \frac{\epsilon}{2\tau k} \right)^{\frac{3}{2}}~
    (1+i\tau d^2)^{-k} - \exp\left( -\frac{\tau kd^2}{\epsilon}\right)
    \exp\left( -i \left( \tau kd^2-\frac{3\pi}{4}\right)\right)\right| \\ \\
    \ds ~~~~ \leq \left( \frac{\epsilon}{2\tau k} \right)^{\frac{3}{2}}~
    (1+\tau^2(h^2-1)^2|\xi|^4)^k +
    \exp\left(-\frac{\tau k(h^2-1)|\xi|^2}{\epsilon}\right),
    \end{array}$$
we conclude that the results obtained are equal to the conclusions
obtained in (\ref{17*}). In this sense we can say that
    \begin{eqnarray}
    ||E_{h,\tau}^\ast-R_\tau R_h E||_{l_1(G_h \times
    (T_0,+\infty)_\tau)} & \leq & C_2(\epsilon,h,\tau) \label{24*}
    \end{eqnarray}

where $C_2(\epsilon,h,\tau)$ is a quantity that tends to zero when
$\epsilon<\tau^8$ and $\ds \frac{\tau}{h^2}<\frac{1}{6\pi^2}$. From
(\ref{22*}) and (\ref{24*}) we obtain
    \begin{eqnarray*}
    & ||E_{h,\tau}^\ast-R_\tau R_h E||_{l_1(G_h \times
    (T_0,+\infty))} \\ & \\
    \leq &
    \left\{ \begin{array}{ccc}
              T_0 + Vol(G_h)~\frac{T_0}{(4\pi)^{\frac{3}{2}}} + C_2(h,h,\tau) & \mbox{if} & \tau \geq 1 \wedge
              \epsilon<\tau^8 \wedge \frac{\tau}{h^2}<\frac{1}{6\pi^2} \\
              &  &  \\
              T_0 + Vol(G_h)~\frac{T_0+m_0^2}{(4\pi)^{\frac{3}{2}}} + C_2(h,h,\tau) & \mbox{if} &
              0 < \tau < 1 \wedge \epsilon<\tau^8 \wedge \frac{\tau}{h^2}<\frac{1}{6\pi^2} \\
            \end{array}
    \right.. \label{25*}
    \end{eqnarray*}

In this conditions we can present the following theorem
    \begin{theo}\label{convteo2}
    For $h \rightarrow 0$ and $\tau \rightarrow 0$ we have the
    convergence
        \begin{eqnarray*}
        ||E_{h,\tau}^\ast(\cdot,t)-R_\tau R_h E(\cdot,t)||_{l_1(G_h \times
        \BR_\tau^+)} \rightarrow 0.
        \end{eqnarray*}
    \end{theo}
    \begin{proof}
    For an arbitrary $\delta$ we can choose $h$ and $\tau$ such that
        \begin{eqnarray*}
        ||E_{h,\tau}^\ast(\cdot,t)-R_\tau R_h E(\cdot,t)||_{l_1(G_h \times
        \BR_\tau^+)} & \leq & \delta.
        \end{eqnarray*}
    If we consider the same $T_0$ of the Theorem 1 and we use
    $T_0^+$ in (\ref{22*}) and $T_0^-$ for (\ref{24*}) we obtain the
    desired result.
    \end{proof}




{\bf Acknowledgement} {\it The second author wishes to express his
gratitude to {\it Funda\c c\~ao para a Ci\^encia e a Tecnologia} for
the support of his work via the grant {\tt SFRH/BSAB/495/2005}.}


\end{document}